\documentclass[twoside]{article}

\usepackage[accepted]{aistats2023}
\usepackage[round]{natbib}

\usepackage[utf8]{inputenc} 
\usepackage[T1]{fontenc}    
\usepackage{hyperref}       
\usepackage{url}            
\usepackage{booktabs}       
\usepackage{amsfonts}       
\usepackage{nicefrac}       
\usepackage{microtype}      
\usepackage{xcolor}         

\usepackage{graphicx}

\usepackage{stackengine}
\usepackage{comment}
\usepackage{algorithm}
\usepackage{algpseudocode}

\usepackage{amsmath}

\DeclareMathOperator*{\argmin}{arg\,min}
\DeclareMathOperator*{\KL}{KL}
\DeclareMathOperator*{\B}{Bernoulli}
\DeclareMathOperator*{\G}{Gamma}
\DeclareMathOperator*{\Beta}{Beta}

\newtheorem{proposition}{Proposition}

\newcommand{\bbeta}{\boldsymbol{\beta}}
\newcommand{\blambda}{\boldsymbol{\lambda}}
\newcommand{\balpha}{\boldsymbol{\alpha}}
\newcommand{\bsigma}{\boldsymbol{\sigma}}

\usepackage{array}
\newcolumntype{H}{>{\setbox0=\hbox\bgroup}c<{\egroup}@{}}

\begin{document}

%

%

\twocolumn[

\aistatstitle{Transport Elliptical Slice Sampling}

\aistatsauthor
{%
  Alberto Cabezas
   \And
   Christopher Nemeth
}

\aistatsaddress{
Lancaster University \And
Lancaster University } ]

\begin{abstract}
We propose a new framework for efficiently sampling from complex probability distributions using a combination of normalizing flows and elliptical slice sampling \citep{murray2010elliptical}. The central idea is to learn a diffeomorphism, through normalizing flows, that maps the non-Gaussian structure of the target distribution to an approximately Gaussian distribution. We then use the elliptical slice sampler, an efficient and tuning-free Markov chain Monte Carlo (MCMC) algorithm, to sample from the transformed distribution. The samples are then \textit{pulled back} using the inverse normalizing flow, yielding samples that approximate the stationary target distribution of interest. Our transport elliptical slice sampler (TESS) is optimized for modern computer architectures, where its adaptation mechanism utilizes parallel cores to rapidly run multiple Markov chains for a few iterations. Numerical demonstrations show that TESS produces Monte Carlo samples from the target distribution with lower autocorrelation compared to non-transformed samplers, and demonstrates significant improvements in efficiency when compared to gradient-based proposals designed for parallel computer architectures, given a flexible enough diffeomorphism.
\end{abstract}

\begin{figure*}[t]
    \centering
    \includegraphics[width=2\columnwidth]{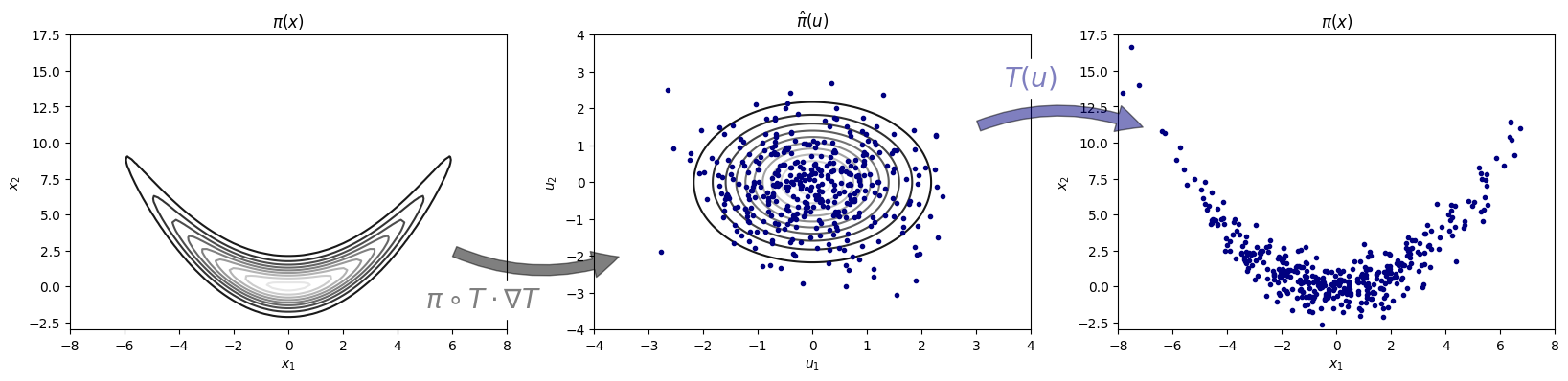}
    \caption{Illustration of Algorithm \ref{tran_ellip} using an exact transport map, i.e. equality in \eqref{transform0} holds: sampling from the Banana density $\pi(x_1, x_2) \propto \exp\left(-[x_1^2/8 + \left(x_2 - x_1^2/4\right)^2]/2\right)$ using the transport map $T(u_1, u_2) = (\sqrt{8}u_1, u_2 + 2u_1^2)$ starts by transforming the target space to the reference space via a change of variables, drawing samples from an ellipsis on the extended reference space (not pictured) and pushing samples back to the target space.}
    \label{fig:first}
\end{figure*}

\section{INTRODUCTION}

Markov Chain Monte Carlo (MCMC) algorithms enable scientists to draw samples from complex distributions, which are typically produced by models that aim to represent the intricate details found in real-world datasets. The exploration of these complex and high-dimensional distributions is challenging, and to be efficient, practitioners use the local pointwise information of the target distribution to create a Markov chain of dependent samples. The ideal outcome would be to have independent samples, but the Markov chain approach generates samples that are correlated sequentially. Therefore, a major focus in MCMC research is to develop algorithms that reduce these correlations and generate samples that approximate independence.

Designing efficient MCMC algorithms usually relies on using local gradient information from the target distribution; by discretizing, for example, Hamiltonian \citep{duane1987hybrid, neal2011mcmc} or Langevin \citep{rossky1978brownian, grenander1994representations} dynamics of a process stationary on our target distribution. Calculating gradients has been automated \citep{linnainmaa1976taylor} but optimizing these algorithms to efficiently minimize both computations and correlations between sequential samples requires algorithmic parameters to be manually tuned. Much work has been dedicated to developing efficient, black-box methods to tune these parameters, with notable examples including the NUTS \citep{hoffman2014no} algorithm which is widely available in probabilistic programming languages \citep{salvatier2016probabilistic, carpenter2017stan, bingham2019pyro, phan2019composable}. 

Within the machine learning community, variational inference \citep[VI;][]{jordan1999introduction} has grown in popularity as an inexact but comparatively faster approach to solving the same inferential problem. As such, MCMC has lost its preferential status as the default approach for Bayesian inference for prediction and uncertainty quantification in this thriving community. Recent efforts \citep{hoffman2021adaptive, hoffman2022tuning} have focused on speeding up MCMC by focusing on widening instead of lengthening computations on modern computer architectures, e.g. utilizing GPUs or TPUs, which allow for vast parallel computations. Tuning parallel MCMC chains has proven to be a somewhat different challenge from its sequential counterpart \citep{radul2020automatically} and parallel efforts need to consider the lockstep necessity of gradient evaluations of parallel chains on modern vector oriented libraries \citep{abadi2016tensorflow, jax2018github, NEURIPS2019_9015}.

\section{TRANSPORT ELLIPTICAL SLICE SAMPLER}

In this paper we assume that $x \in \mathcal{X} \subset \mathbb{R}^d$ are model parameters and $\mathcal{D}$ represents our data. Our goal is to then approximate the posterior distribution, where by Bayes rule the posterior density is given by $\pi(x) \propto L(\mathcal{D}|x)\pi_0(x)$, for $L(\mathcal{D}|x)$ the likelihood function and $\pi_0(x)$ the prior density. Our goal is to introduce a new MCMC algorithm which leverages the tuning-free nature of elliptical slice sampling with the efficient density transformation tools of normalising flows, thus creating the \textit{transport elliptical slice sampler} (TESS); an adaptive mechanism that allows scientists to perform fast parallel sampling from unnormalized densities. An intuitive pictorial representation of our TESS algorithm is given in Figure \ref{fig:first}.


\subsection{Elliptical slice sampling}

Introduced by \citet{murray2010elliptical} as a simple MCMC algorithm with no tuning parameters, the elliptical slice sampler builds on a Metropolis-Hasting sampler introduced by \citet{neal1999}, which is designed for situations where the prior $\pi_0(x)$ is Gaussian. Without loss of generality, we can assume that the prior is a standard Gaussian density $\pi_0(x) = \phi(x)$\footnote{shift and scale $x$ if non-standard.}. The algorithm of \citet{neal1999} proceeds by first proposing new state of the Markov chain, $x' = \sqrt{1 - \beta^2}x + \beta v,$ where $v \in \mathbb{R}^d$ is an independent \textit{momentum} variable following a standard Gaussian distribution. The proposal moves $x$ along the half ellipse which connects the points $v$ and $-v$ which pass thorough $x$, for values $\beta \in [-1, 1]$. 

Elliptical slice sampling, instead, uses the proposal $x' = x \cos\theta + v \sin\theta$ which moves on the full ellipse connecting $x$, $-x$, $v$ and $-v$ for $\theta \in [0, 2\pi]$. Both proposals leave the prior density invariant and elliptical slice sampling uses the slice sampling algorithm \citep{neal2003slice} to choose a value $\theta$ which ensures that the likelihood $L(\mathcal{D}|x)$ is invariant. Overall, this proposal scheme keeps the target posterior invariant \citep{murray2010elliptical}, details of which are presented in the Supplementary material for completeness.


\subsection{Normalizing flows}


Normalizing flows \citep[NF;][]{rezende2015variational} are a flexible class of transformations produced by the sequential composition of invertible and differentiable mappings. Using NF involves choosing a simple \textit{reference density}, for example a standard Gaussian distribution $\phi(\cdot)$, and a  parameterized diffeomorphism $T_\psi$, with optimized parameters $\psi$, to transform the reference density to our target $\pi(x)$ via a change of variables. In other words, we want to find a map $T_\psi$ such that for $u \sim \phi$ and $x = T_\psi(u)$ we have $x \sim \pi$. Assuming this function exists, applying a change of variable yields the following identities
\begin{align}
    \pi(x) &= \phi(T_\psi^{-1}(x))|\det \nabla T_\psi^{-1}(x)| =: \hat{\phi}(x) \label{transform-1}\\
    \phi(u) &= \pi(T_\psi(u)) |\det \nabla T_\psi(u) | =: \hat{\pi}(u) , \label{transform0}
\end{align}
where $\nabla T_\psi$ and $\nabla T_\psi^{-1}$ are the Jacobian matrices of $T_\psi$ and its inverse, respectively. In the context of VI, an approximation of $\hat{\phi}(x)$ would serve as an approximation to our target density when carrying-out inference, since this approximation is both normalized and trivial to sample from.

\subsection{Fixed transport maps with elliptical slice sampling}

To fulfill our requirement for a simple and cost-effective MCMC proposal, we begin by generalizing the dimension-independent, gradient-free, and tuning-free elliptical slice sampler. We will add tuning parameters to our generalized elliptical slice sampler using NF. The diffeomorphism $T_\psi$ will be responsible for efficiently exploring the posterior target density by transforming the proposal's dynamics and tracing the contours of a standard Gaussian density to follow the contours of an approximation of the target. Following previous works in the transport Monte Carlo field (as described in Section \ref{sec:related_work}), we present TESS as a two-step procedure. Firstly, we learn the transport map between the target and reference densities. Secondly, we utilize the transport map within the elliptical slice sampler to generate samples from the target density.

\textbf{1. Map optimization}
To estimate the parameters $\psi$ of our NF map we minimize a divergence between our target density $\pi(x)$ and the \textit{push-forward} reference density $\hat{\phi}(x)$ \eqref{transform-1}. For our intended purpose, by the law of the unconscious statistician, this is equivalent to minimizing the divergence between the \textit{pull-back} target density $\hat{\pi}(u)$ \eqref{transform0} and the reference density $\phi(u)$. The Kullback-Leibler divergence \citep[KL;][]{kullback1951information} is arguably the most widely used and studied divergence, here presented in the context of approximate Bayesian inference but also studied in other branches of statistics and information theory \citep{Joyce2011}. It not only has a tractable Monte Carlo estimate, it is directly related to the foundation of VI and provides intuition into the connection between maximizing the likelihood of observational data and minimizing the distance between target and reference densities \citep{blei2017variational},
\begin{align}
    \KL(\pi || \hat{\phi}) = \int \log\frac{\pi(x)}{\hat{\phi}(x)}\pi(x) dx. \label{KLD}
\end{align}

The optimal transport map is found by optimizing the the parameters $\psi$ of the diffeomorphism $T_\psi$ such that the Kullback-Leibler divergence between the target and reference densities is minimised, i.e. 
\begin{align}
    \psi^* = \argmin_{\psi \in \Psi} \KL(\pi||\hat{\phi}). \label{argmin}
\end{align}

\textbf{2. Sampling from the target} Our proposed sampling method generalizes the elliptical slice sampler by targeting the extended state space $\pi(x)\phi(v)$ for any posterior density $\pi(x)$, regardless of the choice of prior distribution. The target density is preconditioned using a transform via a normalizing flow to map to a standard Gaussian distribution. That is, given a map $T_{\psi^*}$, with fixed parameters $\psi^*$, such that $\hat{\pi}(u) \approx \phi(u)$ we proceed as follows: i) from an initial state $(x, v) = (T_\psi^*(u), v)$, ii) move around an ellipse connecting $u$ and $v$ and iii) accept the new state according to a slice variable chosen uniformly on the interval $[0,\hat{\pi}(u)\phi(v)]$. One iteration of this method is detailed in Algorithm \ref{tran_ellip}. 

\begin{algorithm} 
\caption{Transport Elliptical Slice Sampler} \label{tran_ellip}
\begin{algorithmic}[1]
\Require $u, T_\psi(\cdot), \hat{\pi}(\cdot)$
\State $v \sim \mathcal{N}(0, \mathbb{I}_d)$ \label{v}
\State $w \sim \text{Uniform}(0, 1)$ \label{w}
\State $\log s \gets \log \hat{\pi}(u) + \log \phi(v) + \log w$ \label{y}
\State $\theta \sim \text{Uniform}(0, 2\pi)$ \label{theta}
\State $[\theta_{min}, \theta_{max}] \gets [\theta-2\pi, \theta]$ \label{theta_range0}
\State $u' \gets u \cos\theta + v \sin\theta$ \label{u'}
\State $v' \gets v \cos\theta - u \sin\theta$ \label{v'}
\If{$\log \hat{\pi}(u') + \log \phi(v') > \log s$} \label{acc}
    \State $x' \gets T_\psi(u')$ \label{x'}
    \State \textbf{Return} $x', u'$
\Else
    \If{$\theta < 0$}
        \State $\theta_{min} \gets \theta$
    \Else
        \State $\theta_{max} \gets \theta$
    \EndIf
    \State $\theta \sim \text{Uniform}(\theta_{min}, \theta_{max})$ \label{theta2}
    \State Go to \ref{u'}.
\EndIf
\end{algorithmic}
\end{algorithm}

\begin{proposition} \label{invariance}
The transition kernel of the Markov chain derived from Algorithm \ref{tran_ellip} leaves the target density $\pi(x)\phi(v)$ invariant.
\end{proposition}

The TESS algorithm is likely to be geometrically ergodic under certain transformations, if those transformations lead to nice tail properties on the \textit{pulled back} target. A sketch of this argument follows from three key components: (i) \cite{natarovskii2021geometric} show that the standard elliptical slice sampler is geometrically ergodic if the target density has tails which are rotationally invariant and monotonically decreasing, e.g. $\exp(-c||x||)$ for $c>0$. (ii) This implies geometric ergodicity for TESS if for a target density $\pi$ and Markov transition kernel $P(x, \cdot)$, with $C>0$ and $\gamma \in (0, 1)$, geometric ergodicity of the elliptical slice sampler holds when
\begin{align*}
    ||P^n(x, \cdot) - \pi||_{TV} \leq C(1 + ||x||)\gamma^n, \quad \forall n \in \mathbb{N}, \forall x \in \mathbb{R}^d.
\end{align*}
Then for $\tilde{P}(x, \cdot)$ the transition kernel of TESS we have,
\begin{align*}
|| \tilde{P}^n(u, \cdot) - \hat{\pi}||_{TV} &= || P^n(T_\psi(u), \cdot) - \hat{\pi}||_{TV} \nonumber \\ 
&\leq C(1 + ||T_\psi(u)||)\gamma^n,
\end{align*}
which holds only if the transformation $T_\psi$ leads to nice tail properties for $\hat{\pi}$. (iii) Following from Theorems 2 and 3 of \cite{johnson2012variable}, if there exists a diffeomorphism which ensures that $T$ pulls in the tails of the distribution enough, then geometric ergodicity holds on the transformed distribution. An open question is to determine the necessary conditions on $T_\psi$ for this result to hold beyond simple transformations.

\subsection{Adaptive transport maps}

There are two key components to TESS, the MCMC sampling phase and the transformation function $T_\psi$, which so far we have treated as two independent procedures. However, the function $T_\psi$ is parameterised by $\psi$ and these parameters must be learnt using samples from the target $\pi(x)$. We therefore propose an adaptive version of TESS which alternates between optimizing $\psi$ and sampling $x$ to produce an accurate map between the reference measure and the target distribution. 

The parameters $\psi$ are optimized by first running the TESS sampling procedure (Alg. \ref{tran_ellip}) using $k$ parallel Monte Carlo chains with an initial value of $\psi$, resulting in $k$ approximate samples from our target $\pi(x)$. We then run $m$ iterations of a stochastic gradient descent algorithm on the loss function 
\begin{align}
    \KL(\pi(x)||\hat{\phi}(x)) &\approx \frac{1}{k}\sum_{i=1}^k \log \frac{\pi(x_i)}{\hat{\phi}(x_i)}. \label{reverse_kld_}
\end{align}
The warm-up stage of the sampler repeats this process for $h$ epochs with batches of size $k$, adjusting the inherited parameters from the previous epoch and finally fixing the parameters to then iterate $N$ times Algorithm \ref{tran_ellip}, generating samples from our extended target space $\pi(x)\phi(v)$. This adaptive sampling algorithm is detailed in Algorithm \ref{adapt_trans_ellip}.

An important property of the Kullback-Leibler divergence is that it is an asymmetric divergence, i.e. $\KL(\pi||\hat{\phi}) \neq \KL(\hat{\phi}||\pi)$. Minimizing $\KL(\hat{\phi}||\pi)$ forces $\pi(x)$ to cover the mass of $\hat{\phi}(x)$ thus produces a poor approximation of the tails of the posterior target density. Alternatively, minimizing $\KL(\pi||\hat{\phi})$ forces $\hat{\phi}(x)$ to cover the mass of $\pi(x)$, providing an overconfident approximation to the target density that can be corrected using a sampling method that leaves the target distribution invariant.

\begin{algorithm} 
\caption{Adaptive TESS} \label{adapt_trans_ellip}
\begin{algorithmic}[1]
\Require $u^{(0)}_{1:k}, h, m, N, \texttt{TESS}$ \Comment{\texttt{TESS} applies Algorithm \ref{tran_ellip}}
\State Set initial parameters of $T_\psi$ and $\hat{\pi}$.
\For{$t \gets 1,\dots, h$} \Comment{Warm-up}
\For{$i \gets 1, \dots, k$}
\State $x^{(t)}_i, u^{(t)}_i \gets \texttt{TESS}(u^{(t-1)}_i, T_\psi, \hat{\pi})$
\EndFor
\State Update $\psi$ in $T_\psi$ by running $m$ iterations of gradient descent on \eqref{reverse_kld_} using samples $x^{(t)}_{1:k}$.
\EndFor
\State $u^{(0)}_{1:k} \gets u^{(h)}_{1:k}$
\For{$t \gets 1, \dots, N$} \Comment{Sampling}
\For{$i \gets 1, \dots, k$}
\State $x^{(t)}_i, u^{(t)}_i \gets \texttt{TESS}(u^{(t-1)}_i, T_\psi, \hat{\pi})$
\EndFor
\EndFor
\State Return $x^{(1)}_{1:k}, \dots, x^{(N)}_{1:k}$
\end{algorithmic}
\end{algorithm}

We follow the approach of \cite{hoffman2019neutra} and initialize the parameters of the NF using an approximation of the parameters that minimize $\KL(\hat{\phi}||\pi)$ via a stochastic gradient descent scheme. In other words, minimizing the Monte Carlo approximation 
\begin{align}
    \KL(\phi(u)||\hat{\pi}(u)) &\approx \frac{1}{M}\sum_{i=1}^M\log\frac{\phi(u_i)}{\hat{\pi}(u_i)}, \quad u_i \overset{iid}{\sim}\phi. \label{forward_kld_}
\end{align}

\subsection{Choice of transport map}
\label{sec:choice_map}

There is a wide class of linear and nonlinear functions which can be used within our normalizing flow map. In this paper, we focus on the coupling architecture for $T_\psi$ introduced by \citet{dinh2014nice}. Consider the disjoint partition $x = (x^A, x^B) \in \mathbb{R}^p \times \mathbb{R}^{d-p}$ and a coupling function $t(\cdot; \psi): \mathbb{R}^p \rightarrow \mathbb{R}^p$ parameterized by some set of parameters $\psi$. Then, one can define a transformation $G: \mathbb{R}^{d} \rightarrow \mathbb{R}^{d}$ by the formula
\begin{align}
    x^A &= t(u^A; \Psi(u^B)) := e^{\psi_1} \odot u^A + \psi_2 \label{affine} \\
    x^B &= u^B, \label{const}
\end{align}
given parameters $\Psi: \mathbb{R}^{d-p} \rightarrow \mathbb{R}^{p} \times \mathbb{R}^{p}$ learned only from the extended input. Here we assume an affine bijection, defined in \eqref{affine}, and make $\Psi$ a dense feedforward neural network, for further generalizations and variations see \citet{kobyzev2020normalizing}. The main practical advantages of the coupling architecture with affine transformations are that it is easily inverted through a shift and scale of the transformed $x^A$ with parameters given by the unchanged $x^B=u^B$, and that the modulus determinant of its Jacobian matrix can be easily computed as $|\det \nabla G(x) | = \prod_{i=1}^d (e^{\psi_1})_i$. Furthermore, since the inverse of the transformation is of similar structure, also its constant of volume change can be easily derived as $|\det \nabla G^{-1}(x)| = \prod_{i=1}^d (e^{-\psi_1})_i$. Both of these are using parameters given by $\Psi(u^B) = \Psi(x^B)$ and we drop the absolute value from our computations since the values being multiplied are non-negative. We allow for arbitrary complexity of our NF by introducing a transformation $D: \mathbb{R}^{d} \rightarrow \mathbb{R}^{d}$ with the same structure as $G$ but with the roles of the random variables reversed, i.e. $x^A = u^A$ and $x^B = t(u^B; \Psi(u^A))$. Hence making our final NF a sequential composition of $n \geq 1$ transformations $T_\psi = D_n \circ G_n \circ \cdots \circ D_1 \circ G_1$.


\section{RELATED WORK}
\label{sec:related_work}

\textbf{Elliptical slice sampling} The original elliptical slice sampler paper \citep{murray2010elliptical} presented a simple algorithm that worked well on scenarios of strong prior (Gaussian) information. \cite{nishihara2014parallel} were the first to explore the idea of generalizing this algorithm to any target distribution, while trying to maintain a simple kernel. Their proposal used a Student-t distribution to approximate the target, under the premise that this proposal would adequately cover the tails of the target density. Their work also considered an adaptive mechanism using parallel computing architectures, which accelerated the MCMC sampler by utilizing multiple chains with fewer iterations per chain. \cite{fagan2016elliptical} also used a generalized elliptical slice sampling proposal paired with a preconditioning step to alleviate complex geometry on their target, in their case using expectation propagation to learn correlation structures for subsets of the parameter space. The main difference between previous elliptical slice sampling work and our methodology is the use of normalizing flows to create a transport map between a Gaussian density (for which the sampler works well) and the target density of interest. As shown in Section \ref{sec:experiments}, utilizing the richness of nonlinear transport maps produces a fast and highly efficient MCMC algorithm. 

\textbf{Monte Carlo transport maps} Our work draws inspiration and is closely related to several threads of work that approach the problem of simulation by simplifying the structure of the target density through a preconditioning step. For general MCMC proposals, \cite{parno2018transport} introduced the idea of learning a diffeomorphism using samples from an MCMC algorithm to approximate \eqref{KLD}. Their work built on \cite{el2012bayesian}'s proposal for approximate inference, adding an MCMC kernel that corrects the approximation and provides asymptotically exact samples. Their work showed that a relatively simple transformation can provide valuable information about the global structure of the target density, thus improving the efficiency of MCMC algorithms that use local gradient information on certain, especially degenerate, test cases.

\textbf{MCMC with normalizing flows} The NeuTra Hamiltonian Monte Carlo (HMC) algorithm introduced in \cite{hoffman2019neutra} combines neural transport maps with the HMC algorithm. This builds on the earlier work of \cite{marzouk2016introduction}, who frame the approximate inference problem as solving a two-step process, where firstly an optimization problem is solved to find a preconditioned diffeomorphism which minimizes \eqref{forward_kld_}, and then the preconditioned target is sampled from using a HMC algorithm. The NeuTra algorithm relies on gradient-based proposals to explore the target density. A key difference from the transport elliptical slice sampler is that gradients of the target density are not required, this makes the algorithm faster than gradient-based MCMC algorithms, and as illustrated in Section \ref{sec:experiments},  this is achieved without sacrificing sampling accuracy due to the transport mapping. Additionally, TESS can be applied in settings where it is either infeasible to calculate target gradients, or they may be unstable (e.g the Neal's funnel density \citep{gorinova2020automatic}).  

\section{EXPERIMENTS} \label{sec:experiments} \let\thefootnote\relax\footnotetext{\url{https://github.com/albcab/TESS}}

In this section we compare the performance of the adaptive form of TESS (Alg. \ref{adapt_trans_ellip}) with the performance of several state-of-the-art MCMC algorithms designed for parallel computer architectures. Specifically, MEADS \citep{hoffman2022tuning}, 
ChEES-HMC \citep{hoffman2021adaptive}, 
and the popular NUTS algorithm \citep{hoffman2014no} where an adaptive step size is tuned such that the average cross-chain harmonic-mean acceptance rate is approximately 0.8. We also precondition the latter NUTS method, using the same NF as in TESS, which leads to the NeuTra algorithm \citep{hoffman2019neutra}. We compare the effect of TESS's overfitted adapted transformation against an underfitted transformation (i.e. reversing the KL) which, unlike TESS, is done independently and a priori to the sampling process. 
Each experiment runs all algorithms on 128 parallel chains for 400 warm-up iterations per chain, during which hyperparameters are tuned, followed by 100 iterations used to produce posterior samples, with fixed hyperparameters. MEADS separates the 128 chains into 4 batches of 32 chains each and tunes parameters during all 500 iterations but only the last 100 are used as posterior samples.

The transform map used in all experiments uses $n=2$ transformations of a $\psi$-parameterized dense feedforward neural network with two hidden layers of the same dimension as the input (see Sec. \ref{sec:choice_map} for details). 
The Adam \citep{kingma2014adam} method is used to estimate $\psi$, with a learning rate that decays exponentially over 400 iterations at a rate of 0.1 using a different initial learning rate for each experiment. For the adaptive TESS algorithm, we set $m=1$ on all experiments. 

We compare the experimental results of each algorithm based on their Monte Carlo sample efficiency, as indicated by the maximum integrated autocorrelation time ($\tau_{max}$) with standard deviation ($\sigma_\tau$). Additionally, we present the effective sample size (ESS) in terms of the median worst case integrated autocorrelation time, both for individual chains and when all chains are grouped together. A more efficient algorithm is indicated by lower autocorrelations and higher ESS, as this indicates that samples are closer to being independent. To demonstrate the impact of computational cost on each algorithm, we normalize the ESS by the run time in seconds. ESS/sec considers the time spent adapting and sampling, therefore provides a fair comparison between algorithms. To assess the accuracy of the posterior approximation for each algorithm, we use the kernelized Stein discrepancy with U- and V-statistics, as described in \citep{gorham2017measuring}. Lower values of U- and V- statistics indicate a better approximation of the target posterior. Further information on these diagnostics can be found in the Supplementary Material.

\subsection{Biochemical oxygen demand model} \label{bio}

\begin{table*}[t]
\centering
\begin{tabular}{l|rr|rcrH|rr}
  \hline
Algorithm & $\tau_{max}$ & $\sigma_\tau$ & ESS & ESS/chain & ESS/sec & i & Stein U-stat. & Stein V-stat.\\
   \hline
TESS & \textbf{0.555} & 1.485 & \textbf{11523} & \textbf{90} & \textbf{1129.199} & \textbf{8.822} & \textbf{4.269e+02} & \textbf{4.570e+02}\\
MEADS & 9.959 & 1.468 & 643 & 5 & 208.613 & 1.630 & 1.476e+15 & 1.486e+15\\
ChEES-HMC & 6.406 & 2.228 & 999 & 8 & 224.290 & 1.752 & 1.505e+16 & 1.510e+16\\
NUTS & 9.579 & 1.427 & 668 & 5 & 19.625 & 0.153 & 1.187e+15 & 1.192e+15\\
NeuTra & 9.553 & 1.502 & 670 & 5 & 15.579 & 0.121 & 1.082e+15 & 1.087e+15 \\
\hline
\end{tabular}
\caption{Biochemical oxygen demand model. Algorithm diagnostics where $\tau_{\max}$ is the maximum integrated autocorrelation time over all dimensions; ESS is the corresponding minimum effective sample size. Results are averaged over multiple chains of each sampler, and $\sigma_{\tau}$ is the empirical standard deviation of $\tau_{\max}$ over these runs.}
\label{table:biochem}
\end{table*}

We start with an experiment from \citep{parno2018transport} designed to undermine gradient methods because of its rapidly changing posterior correlation structure, which is challenging for standard samplers to explore. Gradient methods capture local geometry, but the local geometry in this example is not representative of the global geometry of the target and thus provides insufficient information for efficient sampling. On the other hand, the non-linear transformation of the target space with a NF-based approach captures the global, non-Gaussian structure of the target density.

\begin{figure}[t]
    \centering
    \includegraphics[width=\columnwidth]{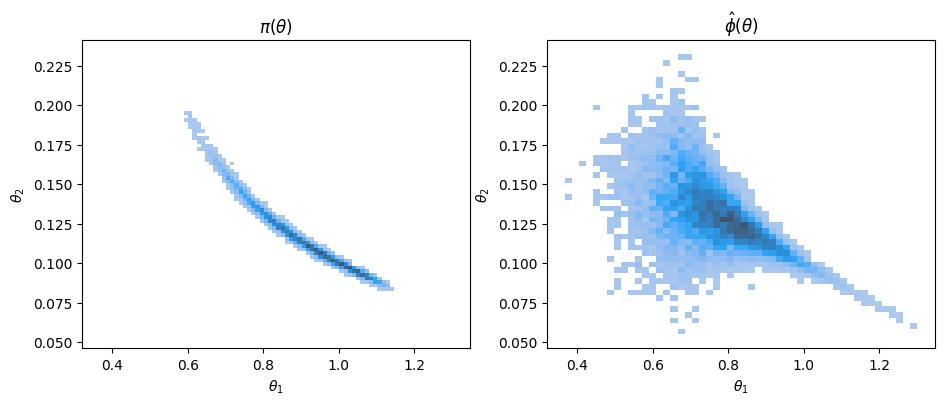}
    \caption{Samples from the target density $\pi(\theta)$ of the Biochemical oxygen demand model acquired by the TESS algorithm, mapped to $\hat{\phi}(\theta)$ \eqref{transform-1}, with diffeomorphism $T_\psi$ learned from the warm-up procedure of Algorithm \ref{adapt_trans_ellip}. With an approximation that overestimates the real variance (right) of our target (left) we are able to capture its global, non-Gaussian structure and explore it using a dimension independent and gradient-free method.}
    \label{fig:tess}
\end{figure}

The simple biochemical oxygen demand model is given by $B(t) = \theta_0(1-\exp(-\theta_1 t))$ for times $t < 5$. In this synthetic data experiment, we set the parameters $\theta_0=1$ and $\theta_1=0.1$ and simulate $y(t_i)$ observations at times $t_i$ evenly spaced in $[0, 5)$ for $i=1,\dots,20$ such that $y(t_i) = \theta_0(1-\exp(-\theta_1 t_i)) + \epsilon_i$, where $\epsilon_i \sim \mathcal{N}(0, \sigma^2_y)$ and fixed $\sigma^2_y=2\times 10 ^ {-4}$. 
The target posterior density is given by the likelihood $L(\mathbf{y}|\theta_0, \theta_1) = \prod_i \mathcal{N}(y(t_i); B(t_i;\theta_0, \theta_1), \sigma^2_y)$ and flat prior $\pi_0(\theta_0, \theta_1) \propto 1$. The numerical results are shown in Table \ref{table:biochem} and Figure \ref{fig:tess} plots the Monte Carlo approximation of the posterior for the original and transformed densities.

It is clear from the results that local gradient information is insufficient to efficiently sample from the rapidly changing local correlation structure of the target density. On the other hand, the learned transport map from the warm-up procedure of TESS provides a mass-covering approximation of the global structure of the target, demonstrated in Figure \ref{fig:tess} by $\hat{\phi}(u)$, which allows the algorithm to move farther away from its initial position, exploring efficiently the entire target space, and  yielding not only shorter autocorrelation times, but also the correct posterior estimates of the parameter space. In this specific case, gradient-based algorithms are forced to take very small steps while still encountering large rejection probabilities, thus being inefficient at producing samples from the posterior.  

\subsection{Sparse logistic regression}

\begin{table*}[t]
\centering
\begin{tabular}{l|rr|rcrH|rr}
  \hline
Algorithm & $\tau_{max}$ & $\sigma_\tau$ & ESS & ESS/chain & ESS/sec & i & Stein U-stat. & Stein V-stat.\\
   \hline
TESS & 5.182 & 0.352 & 1235 & 10 & 34.744 & 0.271 & 1.591e+00 & 1.693e+00\\
MEADS & 7.105 & 0.413 & 901 & 7 & 49.453 & 0.386 & 9.408e-01 & 1.079e+00\\
ChEES-HMC & 5.666 & 0.380 & 1130 & 9 & \textbf{81.588} & \textbf{0.637} & 1.193e+00 & 1.312e+00\\
NUTS & 4.734 & 0.833 & 1352 & 11 & 0.379 & 0.003 & 1.004e+00 & 1.138e+00 \\
NeuTra & \textbf{2.482} & 1.949 & \textbf{2579} & \textbf{20} & 0.401 & 0.003 & \textbf{3.618e-01} & \textbf{4.971e-01}\\
\hline
\end{tabular}
\caption{Sparse logistic regression. Algorithm diagnostics where $\tau_{\max}$ is the maximum integrated autocorrelation time over all dimensions; ESS is the corresponding minimum effective sample size. Results are averaged over multiple chains of each sampler, and $\sigma_{\tau}$ is the empirical standard deviation of $\tau_{\max}$ over these runs.}
\label{table:logistic}
\end{table*}


Next, we consider a sparse logistic regression model with hierarchies. Regression parameters of the logistic likelihood are given a horseshoe prior \citep{carvalho2009handling} which induces sparsity on the regressors, i.e. variable selection. These types of hierarchies on the prior scale of a parameter create funnel geometries that are hard to efficiently explore without local or global structure of the target. 

Algorithms are run on the non-centered parametrization \citep{papaspiliopoulos2007general} of our model using the numerical version of the German credit dataset\footnote{\url{https://archive.ics.uci.edu/ml/datasets/statlog+\\(german+credit+data)}}. The target posterior is defined by the likelihood $L(\mathbf{y}|\bbeta, \blambda, \tau) = \prod_i \B(y_i;\sigma((\tau \blambda \odot \bbeta)^T X_i))$, with sigmoid function $\sigma(\cdot)$, and prior $\pi_0(\bbeta, \blambda, \tau) = \G(\tau;1/2, 1/2)\prod_j \mathcal{N}(\beta_j;0, 1)\G(\lambda_j;1/2, 1/2)$. Numerical results for each MCMC algorithm are shown in Table \ref{table:logistic}. Notice how NUTS and NeuTra provide the best results but long sampling times reflect their inefficiency when running in parallel: every iteration takes as long as the longest chain takes to iterate. Waiting for all chains to catch up severely slows down sampling time, the same effect can be observed in all experiments.

As the dimension of the parameter space grows ($d=51$ in this example), TESS will require more samples, i.e. more chains, for a low variance estimate of \eqref{forward_kld_}. In addition, a more complicated NF is required to capture the non-Gaussian structure of the high-dimensional target space. When either of these fail, and the diffeomorphism $T$ is unable to capture the structure of the target space, the simple sampling procedure inherited from the elliptical slice sampler will struggle to sample from the target space, even if producing uncorrelated samples. We purposely illustrate the effect of a deficient transformation on a high dimensional problem in order for the practitioner to understand the caveats of our method. Studying ways to lower the variance of \eqref{forward_kld_}, using control variates \citep{lemieux2014control} and similar methods \citep{botev2017variance}, as well as alternative NF schemes is left to future work.

\subsection{Regime switching Hidden Markov model}

\begin{table*}[t]
\centering
\begin{tabular}{l|rr|rcrH|rr}
  \hline
Algorithm & $\tau_{max}$ & $\sigma_\tau$ & ESS & ESS/chain & ESS/sec & i & Stein U-stat. & Stein V-stat.\\
   \hline
TESS & \textbf{0.267} & 0.893 & \textbf{23985} & \textbf{187} & \textbf{985.969} & \textbf{7.703} & 5.120e-02 & 1.301e-01\\
MEADS & 1.382 & 1.197 & 4631 & 36 & 319.949 & 2.500 & 3.066e-01 & 3.867e-01\\
ChEES-HMC & 3.451 & 1.825 & 1855 & 14 & 121.756 & 0.951 & \textbf{-8.203e-03} & \textbf{7.073e-02}\\
NUTS & 0.282 & 0.403 & 22672 & 177 & 182.255 & 1.424 & 2.222e-02 & 1.009e-01 \\
NeuTra & 0.441 & 1.020 & 14530 & 114 & 209.069 & 1.633 & 1.092e-01 & 1.880e-01\\
\hline
\end{tabular}
\caption{Regime switching Hidden Markov model. Algorithm diagnostics where $\tau_{\max}$ is the maximum integrated autocorrelation time over all dimensions; ESS is the corresponding minimum effective sample size. Results are averaged over multiple chains of each sampler, and $\sigma_{\tau}$ is the empirical standard deviation of $\tau_{\max}$ over these runs.}
\label{table:google}
\end{table*}

A important use of inference and uncertainty quantification is on time series data. In this example, we analyze financial time series, specifically the daily difference in log price data of Google's stock, referred to as returns $r_t$, for $t=1,\dots,431$. We shall assume that at any given time $t$ the stock's returns will follow one of two regimes: an independent random walk regime $r_t \sim \mathcal{N}(\alpha_1, \sigma^2_1)$, or an autoregressive regime $r_t \sim \mathcal{N}(\alpha_2 + \rho r_{t-1}, \sigma_2^2)$. We define the two regimes as $s_t\in \{0, 1\}$ and the probability of switching between, or remaining within a regime at time $t$ will depend on the regime at $t-1$, i.e.  $p_{s_{t-1}, s_{t}}$ for $s_{t-1}, s_t \in \{0, 1\}$. The transition probabilities $p_{1,1}$ and $p_{2,2}$, and their  complementary probabilities $p_{1,2} = 1-p_{1,1}$ and $p_{2,1} = 1-p_{2,2}$ are treated as model parameters. Since the regime at any time is unobserved, we instead carry over time the probability of belonging to either regime as $\xi_{1t} + \xi_{2t} = 1$. Finally, we define the initial values, both for returns $r_0$ and the probability of belonging to one of the two regimes $\xi_{10}$. 

The regime switching model is defined by the likelihood
\begin{align}
    L(\mathbf{r}|\balpha, \rho, \bsigma^2, \mathbf{p}, r_0, \xi_{10}) = \prod_t \xi_{1t}\eta_{1t} + (1-\xi_{1t})\eta_{2t}, \\
    \mbox{where} \ \ \xi_{1t} = \frac{\xi_{1t-1}\eta_{1t}}{\xi_{1t-1}\eta_{1t} + (1-\xi_{1t-1})\eta_{2t}}, \nonumber
\end{align}
and $\eta_{jt} = p_{j,1} \mathcal{N}(r_t;\alpha_1, \sigma_1^2) + p_{j,2} \mathcal{N}(r_t; \alpha_2 + \rho r_{t-1}, \sigma_2^2)$ for $j\in\{0, 1\}$. The prior distributions for the parameters are
\begin{align}
    \alpha_1, \alpha_2, r_0 &\sim \mathcal{N}(0, 1), \
    \rho \sim \mathcal{N}^0(1, 0.1), \\
    \sigma_1, \sigma_2 &\sim \mathcal{C}^+(1), \\
    p_{1,1}, p_{2,2} &\sim \Beta(10, 2),  \
    \xi_{10} \sim \Beta(2, 2),
\end{align}
where $\mathcal{N}^0$ indicates a Gaussian distribution which is truncated at zero and $\mathcal{C}^+$ is the half-Cauchy distribution. Numerical results are shown in Table \ref{table:google}.

The marginal unimodality and somewhat independent correlation structure of the parameters makes this posterior distribution easy to sample from, diagnostic results show the best performance for all algorithms with respect to other models. TESS's learned flexible transformation of the target density, allowing it to propose uncorrelated sequential samples, is fundamental for its superior diagnostics. ChEES-HMC outputs the samples with the lowest Stein discrepancy, but since it uses the same step size for all target dimensions it struggles to mix well on the worst-case dimension. On the other hand, a flexible transport map is able to capture the covariance structure of the target, allowing fast mixing even on the worst-case dimension. Pair density plots can be found in the Supplementary Material.

\subsection{Predator-prey system}


We consider a likelihood defined as a solution of an ODE system, specifically, the predator-prey system defined by the Lotka-Volterra equations \citep{goel1971volterra},
\begin{equation}
    \frac{dp}{dt} = \alpha p - \beta pq, \ \ \mbox{and} \ \
    \frac{dq}{dt} = - \gamma q + \delta pq,
\end{equation}
where $p$ and $q$ are the prey and  predator populations, respectively. We can solve the ODE system of equations numerically and account for measurement error by modelling the observations as $\log p_t \sim \mathcal{N}(\log p(t), \sigma_p^2)$ and $\log q_t \sim \mathcal{N}(\log q(t), \sigma_q^2)$ for all $t>0$. Furthermore, $p(0)$ and $q(0)$ are the initial values. Since we cannot analytically solve the system of equations, we approximate its solution using the Runge–Kutta method, adding an approximation error to our likelihood function. Data for the Hudson's Bay historical lynx-hare population\footnote{\url{http://people.whitman.edu/~hundledr/courses/M250F03/LynxHare.txt}} are used as observations in the model. The likelihood is defined as
\begin{align*}
 L(\mathbf{p}, \mathbf{q}|\theta) = \prod_t
     \mathcal{N}\left(   \begin{pmatrix}
    \log p_t \\
    \log q_t
    \end{pmatrix};
    \begin{pmatrix}
    \log p(t) \\
    \log q(t)
    \end{pmatrix},\begin{pmatrix}
    \sigma_p^2 & 0 \\
    0 & \sigma_q^2
\end{pmatrix}\right)
\end{align*}
where $\theta = (\alpha, \beta, \gamma, \delta, \sigma_p^2, \sigma_q^2, p(0), q(0))$ and $\{p(t), q(t)\}_{t>0}$ are approximate solutions to the Lotka-Volterra system of equations initialized at $(p(0), q(0))$. Prior distributions for parameters are
\begin{align}
    \alpha, \gamma \sim \mathcal{N}^0(1, 1/2), \ \ &
    \beta, \delta \sim \mathcal{N}^0(1/20, 1/20), \\
    \log \sigma_p, \log \sigma_q &\sim \mathcal{N}(-1, 1), \\
    \log p(0), \log q(0) &\sim \mathcal{N}(\log 10, 1),
\end{align}
where $\mathcal{N}^0$ is a Gaussian distribution truncated at zero. 

\begin{figure}[h]
    \centering
    \stackunder{\includegraphics[width=.49\columnwidth]{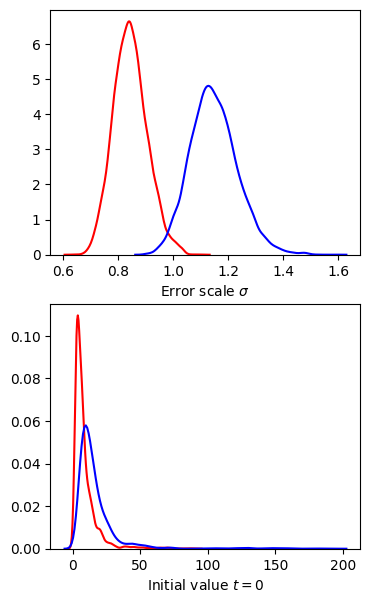}}
    {TESS}
    \stackunder{\includegraphics[width=.478\columnwidth]{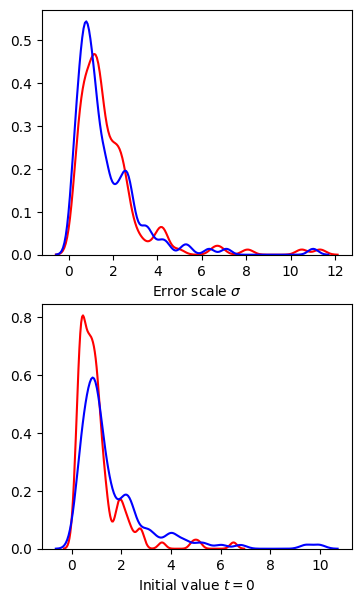}}
    {MEADS}
    \caption{Density plots of the approximate posterior distribution for the initial values and scale parameters from the \textcolor{red}{predator}-\textcolor{blue}{prey} system model, drawn with transport elliptical slice sampling on the left and MEADS on the right.}
    \label{fig:pp}
\end{figure}

This experiment exhibits a situation similar to Section \ref{bio}: gradient methods, without global information on the structure of our target, lack enough information to move efficiently around its rapidly changing correlation structure. On the other hand, TESS captures the global structure of the target using a NF and is able to move purposely around it when sampling. Figure \ref{fig:pp} illustrates the contrast: MEADS is unable to converge towards a sensible solution, exploring a region of the target space with large error variance and insignificant initial positions, both for the predator and the prey populations; on the other hand, TESS is able to converge towards reasonable initial populations and concentrate sampling around small error variance. Samples from the other gradient methods give similar results to MEADS. Gradient methods need a learned correlation matrix that captures the global correlation structure of the target and use gradient information to propose large steps locally, while TESS is able to capture both the global correlation and local structure by learning an overconfident transport map, then using this information on a cheap and gradient-free method for sampling. 

\section{DISCUSSION}

In this paper we proposed TESS, an MCMC algorithm that performs dimension independent and gradient-free sampling from any unnormalized target density. We also proposed an adaptive version of our algorithm that learns a non-Gaussian approximation to the target, helping the algorithm explore complex geometries efficiently. TESS is also able to utilize parallel computer architectures to accelerate sampling from posterior distributions. We believe that this will allow practitioners to perform uncertainty quantification of their models with parallel computational resources and little time.

We found that our algorithm is able to outperform gradient-based competitors in a variety of models. However, it is important to develop flexible transport maps and low-variance Monte Carlo approximations of the KL divergence, specially for high-dimensional models. Future work will explore the role of the transport map on the algorithm's efficiency, its efficacy in capturing issues in Bayesian posterior geometries, and develop flexible transport maps for high-dimensional models.

\subsubsection*{Acknowledgements}
The authors would like to thank the anonymous reviewers for their helpful feedback which has significantly improved the quality of the paper. CN gratefully acknowledges the support of EPSRC grants EP/V022636/1, EP/S00159X/1 and EP/R01860X/1.

\bibliographystyle{unsrtnat}
\bibliography{references}







\appendix
\onecolumn

\section{ELLIPTICAL SLICE SAMPLING ALGORITHM} \label{ellip}

\begin{algorithm} 
\caption{Elliptical slice sampler \citep{murray2010elliptical}} 
\begin{algorithmic}[1]
\Require $x, L(\mathcal{D}|\cdot)$
\State $v \sim \mathcal{N}(0, \mathbb{I}_d)$ \label{p}
\State $w \sim \text{Uniform}(0, 1)$ \label{u1}
\State $\log s \gets \log L(\mathcal{D}|x) + \log w$ \label{u2}
\State $\theta \sim \text{Uniform}(0, 2\pi)$
\State $[\theta_{min}, \theta_{max}] \gets [\theta-2\pi, \theta]$
\State $x' \gets x \cos\theta + v \sin\theta$ \label{q}
\If{$\log L(\mathcal{D}|x') > \log s$} \label{u3}
    \State Return $x'$
\Else
    \If{$\theta < 0$}
        \State $\theta_{min} \gets \theta$
    \Else
        \State $\theta_{max} \gets \theta$
    \EndIf
    \State $\theta \sim \text{Uniform}(\theta_{min}, \theta_{max})$
    \State Go to \ref{q}.
\EndIf
\end{algorithmic}
\end{algorithm}

\section{PROOF}

\subsection{Proof of Proposition~\ref{invariance}}

As established in \cite{murray2010elliptical} and \cite{nishihara2014parallel}, the elliptical slice sampler and generalized elliptical sampler target the correct stationary distribution as the algorithm is reversible and produces an irreducible, aperiodic Markov chain. 

The same result holds for the TESS algorithm from initial state $u=T_{\psi}^{-1}(x)$ and where $(u, v)$ and $(u', v')$ represent the initial and accepted transformed parameters of the sampler (steps \ref{v} and \ref{u'}-\ref{v'}), with $s$ the slice variable (step \ref{y}) and $ \{\theta_k\}_{k=1}^K$ the parameters representing points in the slice expressed in radians until acceptance at $K$ (steps \ref{theta} and \ref{theta2}). Let  
\begin{align}
    \theta'_k = \begin{cases}
    \theta_k - \theta_K, & \text{if } k < K \\
    -\theta_K & \text{if } k = K
    \end{cases},
\end{align}
then by the properties of the elliptical slice sampler, the transformation $(u, v, s,  \{\theta_k\}_{k=1}^K) \mapsto (u', v', s, \{\theta'_k\}_{k=1}^K)$ is bijective, preserves volume and $p(\{\theta_k\}_{k=1}^K | u, v, s) = p(\{\theta'_k\}_{k=1}^K | u', v', s)$. Using the uniform density of the slice variable $s$ it is easy to see that $p(u', v', \{\theta_k\}_{k=1}^K, s|u, v)\hat{\pi}(u)\phi(v) = p(u, v, \{\theta'_k\}_{k=1}^K, s|u', v')\hat{\pi}(u')\phi(v')$, and so if $(u,v) \sim \hat{\pi}(u)\phi(v)$ then $(u',v') \sim \hat{\pi}(u')\phi(v')$. Finally, as $x=T_{\psi}(u)$ we have $(x',v')\sim \pi(T_{\psi}(u'))|\det \nabla T_{\psi}(u')|\phi(v') = \pi(x')\phi(v')$.

\section{DIAGNOSTIC TOOLS CALCULATION DETAILS}

Here we describe the calculation of the maximum integrated autocorrelation time $\tau_{\max}$ and Kernelized Stein discrepancy U- and V- statistics used throughout our results. Assume we have as output from chain $c$ a sequence of $N$ samples from our target $\mathbf{x}_{c,1}, \dots, \mathbf{x}_{c,N}$, where each sample is on a $d-$dimensional parameter space. Then, compute the integrated autocorrelation time for dimension $j=1,\dots,d$ on chain $c$ as
\begin{align}
    \tau_{c,j} &= \frac{1}{2} + 2 \sum_{t=1}^{N-1} \left(1 - \frac{t}{N}\right) \frac{\hat{C}(t)}{2 \hat{C}(0)} \\
    \hat{C}(t) &= \sum_{0<i<N-t}(x_{c,i,j} - \Bar{x}_{c,\cdot,j})(x_{c,i+t,j} - \Bar{x}_{c,\cdot,j}) \\
    \Bar{x}_{c,\cdot,j} &= \frac{1}{N}\sum_{i=1}^N x_{c,i,j}.
\end{align}
The value $\hat{C}(t)$ for all $t=1,\dots,N-1$ is computed by applying the Fourier transform method from \cite{wolff2004monte}. We then define $\tau_{\max}$ and ESS as
\begin{align}
    \tau_{\max} &= \underset{j=1,\dots,d}{\max} \left[\underset{c=1,\dots,C}{\;\text{median}} \; \tau_{c,j} \right] \\
    \text{ESS} &= \underset{j=1,\dots,d}{\min} \left[\underset{c=1,\dots,C}{\;\text{median}} \; \frac{N}{2\tau_{c,j}} \right].
\end{align}

The Kernelized Stein discrepancy's U- and V-statistics are calculated using the inverse multi-quadratic kernel $k(x, x') = (1 + (x - x')^T(x - x'))^{\beta}$ with $\beta=-1/2$ on all experiments as
\begin{align}
    \text{U-stat} &= \frac{1}{CN(CN-1)}\sum_{c,i}\sum_{c'\neq c, i'\neq i}\mathcal{A}_{\pi}\mathcal{A}_\pi'K(\mathbf{x}_{c,i}, \mathbf{x}_{c',i'}) \\
    \text{V-stat} &= \frac{1}{C^2N^2}\sum_{c,i}\sum_{c', i'}\mathcal{A}_{\pi}\mathcal{A}_\pi'K(\mathbf{x}_{c,i}, \mathbf{x}_{c',i'}) \\
    \mathcal{A}_{\pi}\mathcal{A}_\pi'K(x, x') &= \nabla_x \cdot \nabla_{x'} k(x, x') + \nabla_{x}k(x, x') \cdot \nabla_{x'}\log\pi(x') \nonumber \\
    + &\nabla_{x'}k(x, x') \cdot \nabla_{x}\log\pi(x) + k(x, x')\nabla_{x}\log\pi(x) \cdot \nabla_{x'}\log\pi(x).
\end{align}
It can be shown that the U-statistic is an unbiased estimate of $\mathbb{E}_{x, x'\sim \pi^*}[\mathcal{A}_{\pi}\mathcal{A}_\pi'K(x, x')]$ for process $\pi^*$ generating the samples, while the V-statistic is biased but always non-negative \citep{liu2016kernelized}. If $\pi = \pi^*$ then $\mathbb{E}_{x, x'\sim \pi^*}[\mathcal{A}_{\pi}\mathcal{A}_\pi'K(x, x')] = 0$ by Stein's identity \citep{stein2004use}.

\newpage
\section{PLOTS}

\subsection{Regime
switching Hidden Markov model pair plots}

\begin{figure}[h]
    \centering
    \stackunder{\includegraphics[width=.49\columnwidth]{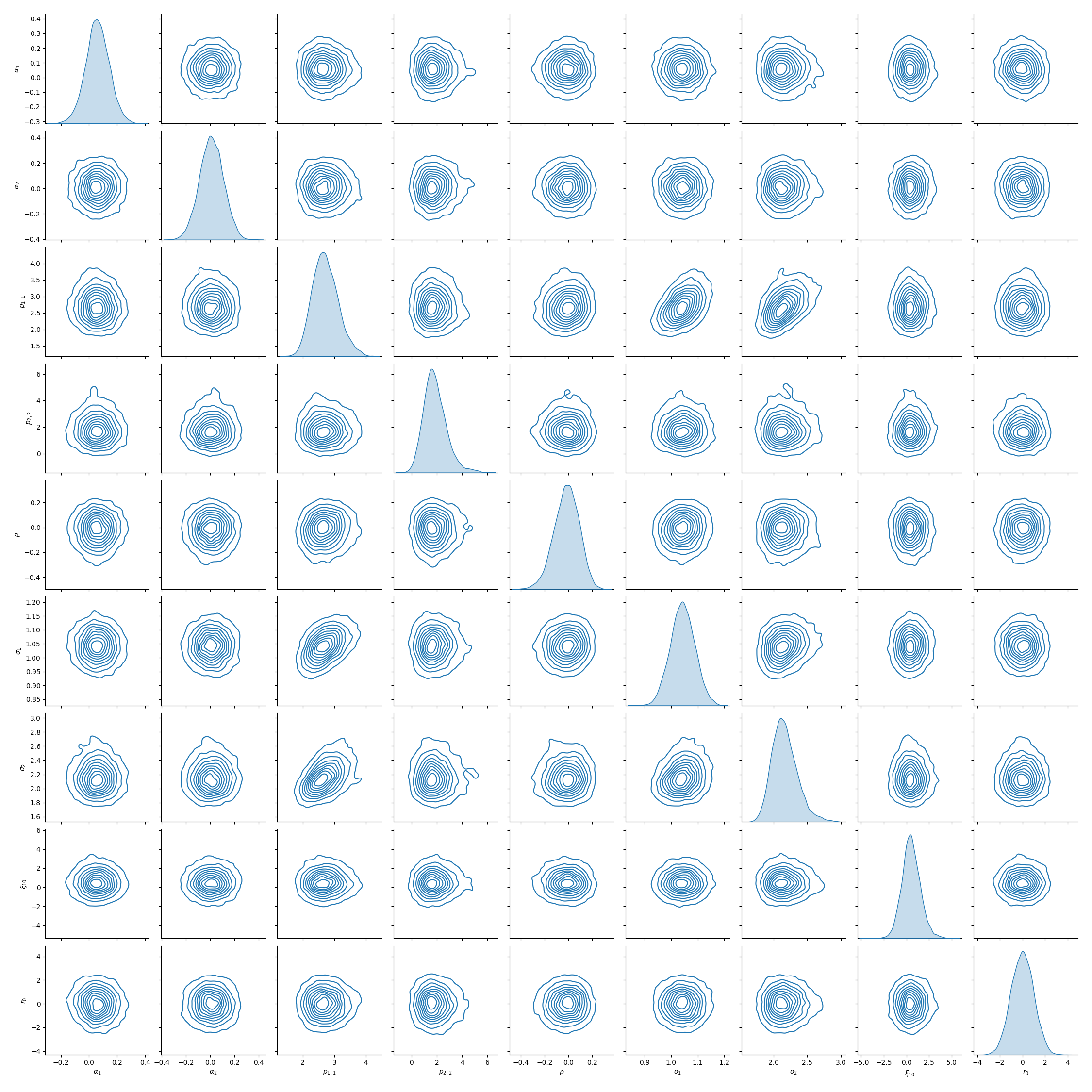}}
    {Samples from TESS}
    \stackunder{\includegraphics[width=.49\columnwidth]{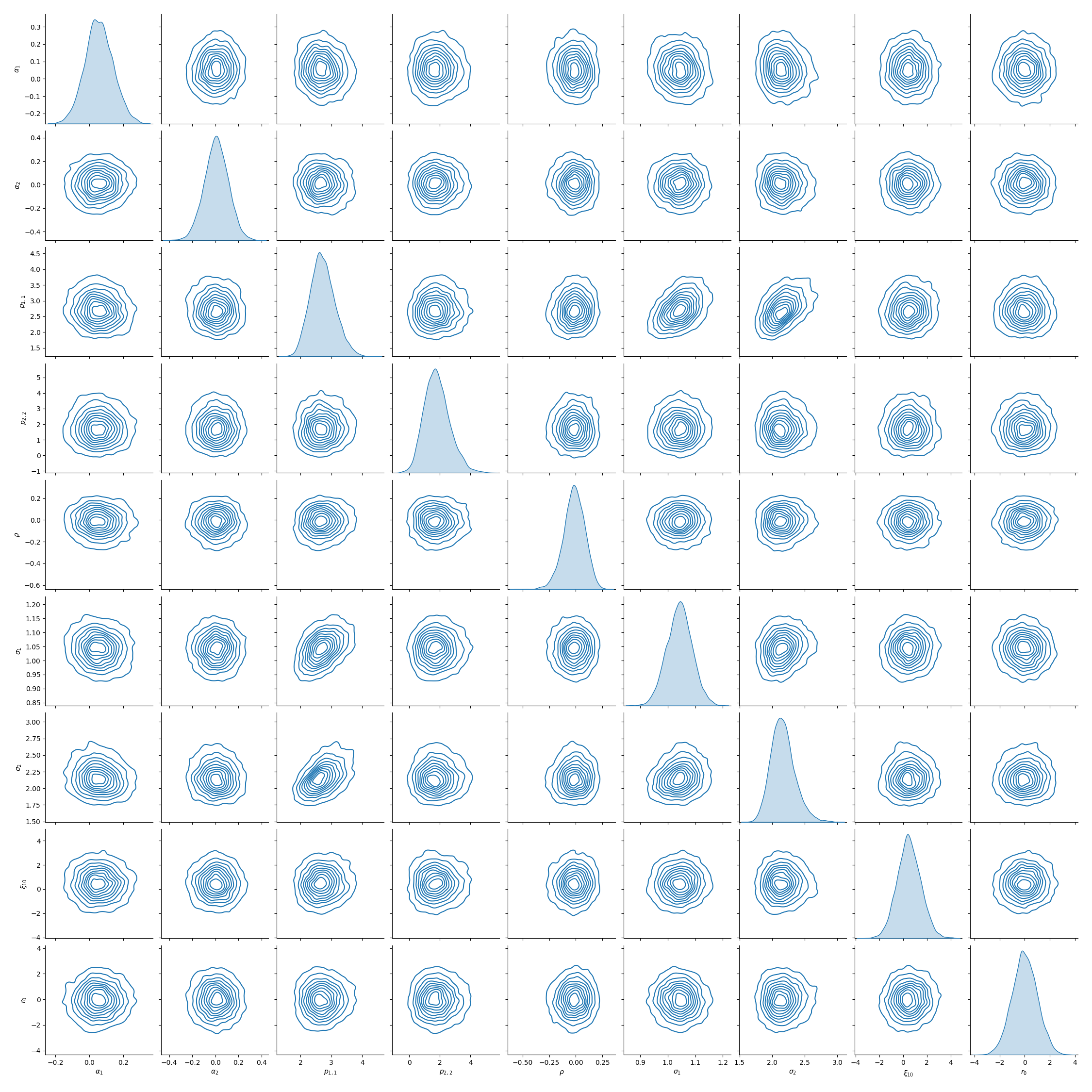}}
    {Samples from MEADS}    
    \caption{Posterior density pair plots for the regime switching hidden Markov model using samples drawn with transport elliptical slice sampling on the left and MEADS \citep{hoffman2022tuning} on the right. Parameters $\rho$, $\sigma_1$ and $\sigma_2$ are log transformed and parameters $p_{1,1}$, $p_{2,2}$ and $\xi_{10}$ are sigmoid function transformed.}
    \label{fig:google}
\end{figure}



\end{document}